\documentclass[aps,prx,reprint,amssymb,superscriptaddress,floatfix]{revtex4-1}
\usepackage{bm}
\usepackage{color}
\usepackage{dcolumn}
\usepackage{amsmath, amsthm}
\usepackage{graphicx}
\usepackage{braket}
\usepackage{everyshi}
\usepackage{hyperref}
\usepackage{units}
\usepackage[euler]{textgreek}

\hypersetup{pdfstartview={FitH},pdfpagemode={UseNone},
    colorlinks,linkcolor=blue, citecolor=blue, urlcolor=blue,
    bookmarks=true, bookmarksopen=true, pdfnewwindow=true}

\begin{document}
  \title{An integrated optical device for frequency conversion across the full telecom C-band spectrum}
  
  \author{Paul Fisher}
  \affiliation{Centre for Quantum Computation and Communication Technology (Australian Research Council),
Centre for Quantum Dynamics, Griffith University, Brisbane, QLD 4111, Australia}
  
  \author{Matteo Villa}
  \affiliation{Centre for Quantum Computation and Communication Technology (Australian Research Council),
Centre for Quantum Dynamics, Griffith University, Brisbane, QLD 4111, Australia}
  
  \author{Francesco Lenzini}
  \affiliation{Centre for Quantum Computation and Communication Technology (Australian Research Council),
Centre for Quantum Dynamics, Griffith University, Brisbane, QLD 4111, Australia}
  \affiliation{Institute of Physics, University of Muenster, 48149 Muenster, Germany}
  
  \author{Mirko Lobino}
  \email{m.lobino@griffith.edu.au}
  \affiliation{Centre for Quantum Computation and Communication Technology (Australian Research Council),
Centre for Quantum Dynamics, Griffith University, Brisbane, QLD 4111, Australia}
  \affiliation{Queensland Micro- and Nanotechnology Centre, Griffith University, Brisbane, QLD 4111, Australia}
  
  \date{\today}
  
  \begin{abstract}
    High-density communication through optical fiber is made possible by Wavelength Division Multiplexing, which is the simultaneous transmission of many discrete signals at different optical frequencies.
    Vast quantities of data may be transmitted without interference using this scheme but flexible routing of these signals requires an electronic middle step, carrying a cost in latency.
    We present a technique for frequency conversion across the entire WDM spectrum with a single device, which removes this latency cost.
    Using an optical waveguide in lithium niobate and two infrared pump beams, we show how to maximize conversion efficiency between arbitrary frequencies by analyzing the role of dispersion in cascaded nonlinear processes.
    The technique is presented generally and may be applied to any suitable nonlinear material or platform, and to classical or quantum optical signals.
  \end{abstract}
  
  \maketitle

\section{Introduction}
Optical communications have greatly increased the data capacity of telecommunication networks since signals can be transmitted faster and with a greater bandwidth compared to copper lines, and can be sent in multiple streams over one fiber with Wavelength Division Multiplexing (WDM). The lowest propagation losses through optical fibre are in the C-band which supports a standard grid of 72 channels spaced by 100 GHz between 1520 nm and 1577 nm \cite{ITU-T2012}. Transferring signals between these channels represents a major speed limitation of current networks since this operation is performed by terminating and measuring all signals with a bank of passive optics and detectors, then retransmitting them using a bank of lasers \cite{Eldada2005}. All-optical signal processing can overcome this speed bottleneck since it could convert data streams between different channels almost instantaneously. In particular, having a single device that can convert data between arbitrary wavelengths without interruption will make a huge impact on current telecommunications technology.
	
There are a number of existing devices for routing WDM signals, known as Reconfigurable Optical Add-Drop Multiplexers (ROADMs). They have been implemented in a number of platforms, including micro-machined reflectors \cite{Pu2000} and thermally-tuned microfabricated resonators \cite{Klein2005}. Although technically flexible, these devices physically route signals without altering the wavelength and are limited to narrow bandwidths or static grid spacing. To switch channels dynamically and leave physical routing to a static structure, uninterrupted channel swapping can be achieved by frequency conversion using nonlinear optics.
  
Nonlinear optical schemes for frequency conversion over the telecom C-band have focused primarily on four-wave mixing in $\chi^{(3)}$ nonlinear materials like single mode \cite{Inoue1994} and photonic crystal fibres \cite{McKinstrie2005,Clark2013}, and silicon waveguides \cite{Li2016}. These schemes suffer from the inherently small value of the $\chi^{(3)}$ coefficient, requiring long fibers and high pump powers, or high quality resonators which reduce the tuning bandwidth. More efficient conversion techniques, sum and difference frequency generation (SFG and DFG respectively) in $\chi^{(2)}$ nonlinear materials such as lithium niobate, have already demonstrated high conversion efficiency in small (5 cm) devices using modest (90 mW CW) pump powers \cite{Langrock2005}. To achieve the frequency shift between two WDM channels however, these two processes must be cascaded.
  
\begin{figure}[h]
\includegraphics[width=\columnwidth]{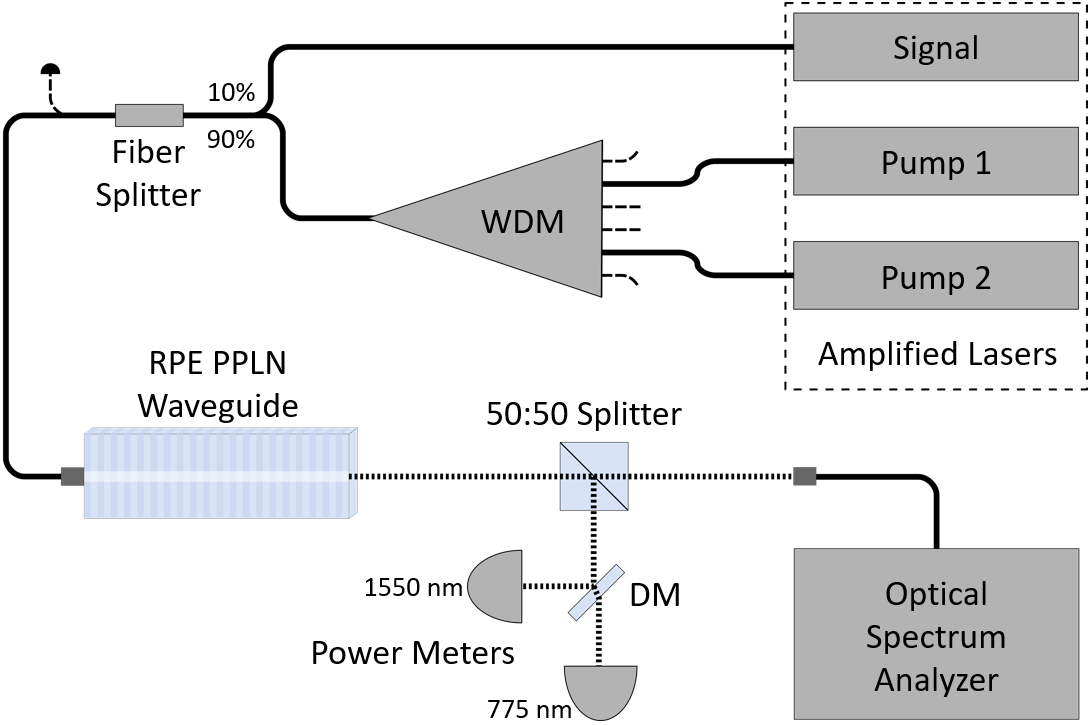}
\caption{\label{fig1}Schematic of the experimental set-up. A WDM module combines the two pump lasers while the signal is added with a 90/10 fiber coupler. All of the components are then coupled to the waveguide.  Light is collected from the waveguide in free space with half diverted to an optical spectrum analyser and the rest collected by two power meters across a dichroic mirror (DM).}
\end{figure}

Cascading can be performed with SFG and DFG steps occurring sequentially, either in two waveguides \cite{Osellame2001} or in opposite directions in the same waveguide \cite{Song2005}. More commonly, cascading refers to performing the SFG and DFG steps simultaneously with two pump lasers, as in Figure \ref{fig1}. Initially, it was proposed as an alternative method to single-step DFG to achieve frequency shifts. Bright pumps generate second harmonic or sum frequency light, which subsequently generates a difference frequency with an input signal. Demonstrations have been performed \cite{Gallo1997,Chou1999} and the technique has been analyzed in several configurations \cite{Gallo1999,Bo2004,Wang2007} and in comparison to single-step DFG \cite{Zhou2003}. Unfortunately, this also results in parametric amplification of the input frequency, which is not suitable for signal dropping.

In an alternative pump configuration, cascading can emulate degenerate four-wave mixing for conversion and signal dropping. In this case, one pump converts the signal by SFG to an intermediate frequency, which is then converted to the target frequency by DFG with a second pump. Experiments have verified this technique in CW and pulsed pump regimes \cite{Jun2003,Min2003}, and have shown its suitability for telecom signals \cite{Furukawa2007}. Subsequent experiments have attempted to improve conversion bandwidth \cite{Lee2004,Lu2010} or enhance signal-dropping and selectivity through engineered poling \cite{Lee2004a} and thermal gradients \cite{Lee2005}. However, these devices have limited operational bandwidth and cannot efficiently convert between any arbitrary pair of channels.
  
Here we propose an optimized protocol for high efficiency frequency conversion across the entire WDM spectrum and demonstrate it using a waveguide fabricated in periodically-poled lithium niobate. We overcome the limitation of previous schemes by analyzing the role of phase mismatch and finding the optimal configuration when using a uniform periodic poling for quasi-phase-matching (QPM). In particular we show that with reverse proton exchanged waveguides in lithium niobate, maximum tunable conversion across the entire telecom C-band can be achieved using pumps around 2.38 $\mu$m. We demonstrate our protocol and show agreement between theory and experiment, performing frequency conversion measurements using pumps also within the C-band. While this is not the pump optimal configuration, we achieve over 30\% conversion and over 80\% signal-dropping in a waveguide with 3.2 cm interaction length.
    
The impact of these results go beyond classical communication and can be applied to quantum networks as well. In particular our device can be used to interface narrow bandwidth Erbium  quantum memory \cite{Rancic2017} with all the WDM channels, greatly increasing the capacity of future quantum repeater networks. Furthermore this protocol may be used where four-wave mixing has been applied in the past, with the enhancement of heralding rate and purity of an SPDC single photon source being an example \cite{Joshi2018}. 

\begin{figure}[t]
\includegraphics[width=\columnwidth]{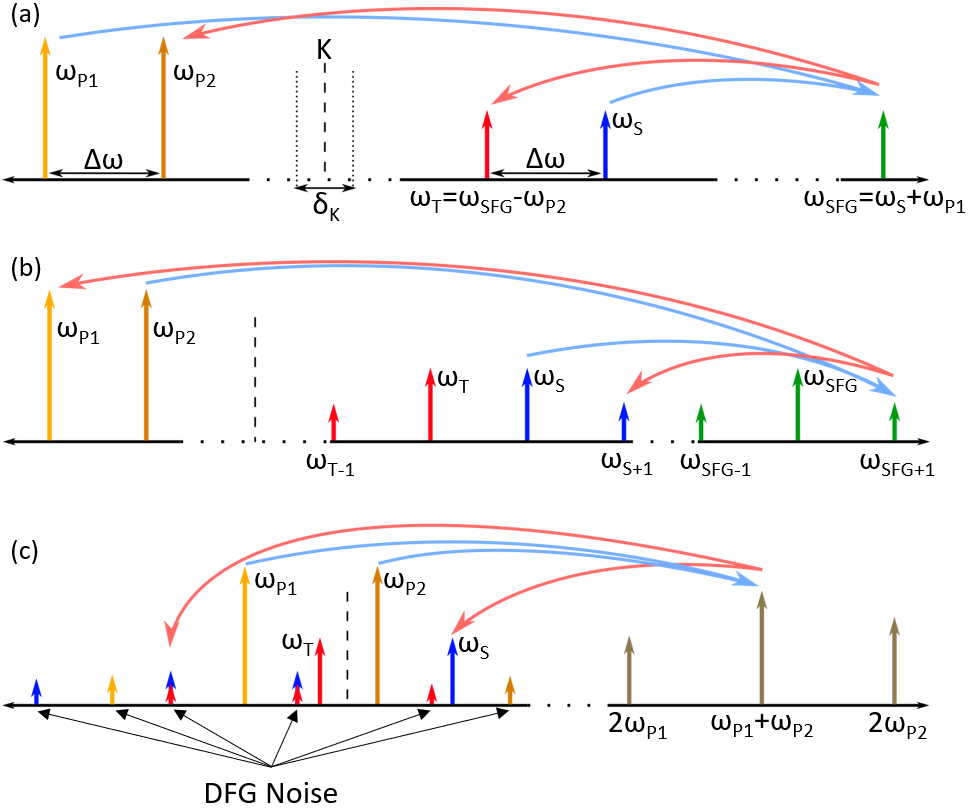}
\caption{\label{fig2} Nonlinear interactions in the waveguide. \textbf{(a)} Correct SFG and DFG interaction for converting light from $\omega_\text{s}$ to $\omega_\text{t}$. First $\omega_\text{SFG}$ is generated by SFG between signal and pump 1. Then the target frequency $\omega_\text{t}$ is generated by DFG $\omega_\text{SFG}-\omega_\text{P2}$. The frequency shift between signal and target is $\Delta\omega=\omega_\text{P1}-\omega_\text{P2}$. \textbf{(b)} Unwanted interaction with the wrong pumps. New frequencies are generated by SFG ($\omega_\text{s(t)}+\omega_\text{P2(P1)}=\omega_\text{SFG+1(SFG-1)}$) and subsequent DFG ($\omega_\text{SFG+1(SFG-1)}-\omega_\text{P1(P2)}=\omega_\text{s+1(t-1)}$)
\textbf{(c)} Sum frequency generation between the two pump beams. This new SFG field also interacts with the signal, target and pumps, and can generate other frequencies by DFG.} 
\end{figure}
\section{Cascaded frequency conversion}
We now derive the conditions for efficiently converting a signal data stream encoded in any of the WDM channels at frequency $\omega_\text{s}$ into a target frequency $\omega_\text{t}$ using a single periodically poled nonlinear waveguide. The first step is SFG between the signal $\omega_\text{s}$ and a pump beam $P_1$ at $\omega_\text{P1}$ which generates a new field at frequency $\omega_\text{SFG}=\omega_\text{P1}+\omega_\text{s}$. The second step is DFG between a second pump $P_2$ at $\omega_\text{P2}$ and the field at $\omega_\text{SFG}$. The frequency of the second pump is chosen such that the DFG goes to the desired target frequency $\omega_\text{t}=\omega_\text{SFG}-\omega_\text{P2}$. Figure~\ref{fig2}a shows a schematic of these interactions.

In this protocol both interactions happen simultaneously in a single nonlinear waveguide where the signal and the two pumps are coupled together. In total, there are 5 frequency components participating in the conversion, each with a different phase velocity. As a consequence, there are two phase mismatches to consider given by
\begin{align}
      \Delta k_\text{SFG}&=k_\text{SFG}-k_\text{s}-k_\text{P1} \\
      \Delta k_\text{DFG}&=k_\text{SFG}-k_\text{t}-k_\text{P2},			
\end{align}
where, for every field involved, $k_i=2\pi n_\text{eff,i}/\lambda_\text{i}$ is the propagation constant of the mode at frequency $\omega_\text{i}=2\pi c/\lambda_\text{i}$, and $n_\text{eff,i}$ is its effective refractive index.
    
In order to derive the conversion efficiency and its bandwidth it is useful to consider the average ($K$) and the difference ($\delta_K$) of the two phase mismatches, 
\begin{align}
	K &= (\Delta k_\text{SFG}+\Delta k_\text{DFG})/2~, \label{K}\\
  \delta_K &= \Delta k_\text{SFG}-\Delta k_\text{DFG}~. \label{dK}
\end{align}    
The value of $K$ determines the optimal poling period $\Lambda=2\pi/K$ since it satisfies the QPM condition $K_\text{QPM}=K-2\pi/\Lambda=0$. It can be shown that for conversion between two arbitrary frequencies the condition $K_\text{QPM}=0$ can be always satisfied with the correct choice of the two pump frequencies $\omega_\text{P1}$ and $\omega_\text{P2}$. The second quantity $\delta_K$ is analogous to the phase-mismatch of the corresponding four-wave mixing process and is the primary factor which limits the overall conversion efficiency.

When the pump powers are optimized, the conversion ($\eta_c$) and signal-dropping ($\eta_d$), that is the fraction of power removed from the signal channel, efficiencies are given by
    \begin{align}
    \eta_c &= \frac{Q^2L^4}{16}\text{sinc}^4\left(\frac{1}{4}\sqrt{\delta_K^2L^2+4QL^2}\right) ~ , \label{etac}\\ 
    \eta_d &= \frac{QL^2}{2}\text{sinc}^2\left(\frac{1}{4}\sqrt{\delta_K^2L^2+4QL^2}\right) - \eta_c ~ . \label{etad}
    \end{align}
$Q$ is a function of the total pump power and is proportional to the square of the $\chi^{(2)}$ nonlinearity.
These efficiencies also depend on the length of the device $L$, illustrating that total phase mismatch $\delta_K L$ is cumulative and becomes more detrimental over longer devices.  Maximum conversion is achieved when $Q = \pi^2/L^2$.  Therefore, longer devices require less pump power to achieve maximum conversion, so the practicalities of available pump powers and damage thresholds must be weighed against tolerance to phase mismatch.

The pump powers are optimised when they are balanced by the ratio
\begin{equation}
\frac{P_1}{P_2} = \frac{\omega_\text{t} n_\text{eff,s} n_\text{eff,P1}A_\text{eff,SFG}}{\omega_s n_\text{eff,t} n_\text{eff,P2}A_\text{eff,DFG}} ~ , \label{PowBal}
\end{equation}
where $A_\text{eff,i}$ are the effective areas of the SFG and DFG processes inside the waveguide. 
This ensures that the processes progress at the same rate as one another and that the only factor unbalancing them is the phase mismatch.  For a detailed derivation of this protocol and all of the introduced quantities and formulae, see supplementary material.

Other factors to be considered in the design of a device relate to the suppression or minimization of unwanted nonlinear processes. Because there are five different frequencies propagating inside the same waveguide, unwanted three-wave mixing interactions are possible. For example the signal and target fields may interact with the wrong pumps, generating new frequencies around $\omega_\text{SFG}$ and subsequent DFG into wrong channels (see Fig.~\ref{fig2}b). Also, the pumps may interact with one another and generate sum frequency or second harmonics, which may then produce other unwanted frequencies in the C-band by DFG (see Fig.~\ref{fig2}c).
    
Unwanted processes can be mitigated or suppressed by ensuring that they are not quasi-phase matched. Because the phase matching bandwidth decreases with increasing interaction length, longer devices reduce these unwanted effects better than shorter ones. The case where the signal interacts with the wrong-pump (Fig.~\ref{fig2}b) and generates the wrong SFG $\bar{\omega}_\text{SFG}=\omega_\text{s}+\omega_\text{P2}$, largely depends on the dispersion properties of the device around $\omega_\text{SFG}$. While the average phase mismatch $\bar{K}$ of this unwanted process is generally non-zero, greater chromatic dispersion around $\omega_\text{SFG}$ means that the magnitude of $\bar{K}$ will be larger and the unwanted process will be less efficient.
Pumps second harmonics $2\omega_\text{P1}$ and $2\omega_\text{P2}$, and SFG $\omega_\text{P1}+\omega_\text{P2}$ are easily reduced by choosing the frequencies $\omega_\text{P1}$ and $\omega_\text{P2}$ away from those of signal and target, but this also reduces the conversion efficiency and tuning bandwidth as the magnitude of $\delta_K$ tends to increase.

\begin{figure}[t]
    \includegraphics[width=\columnwidth]{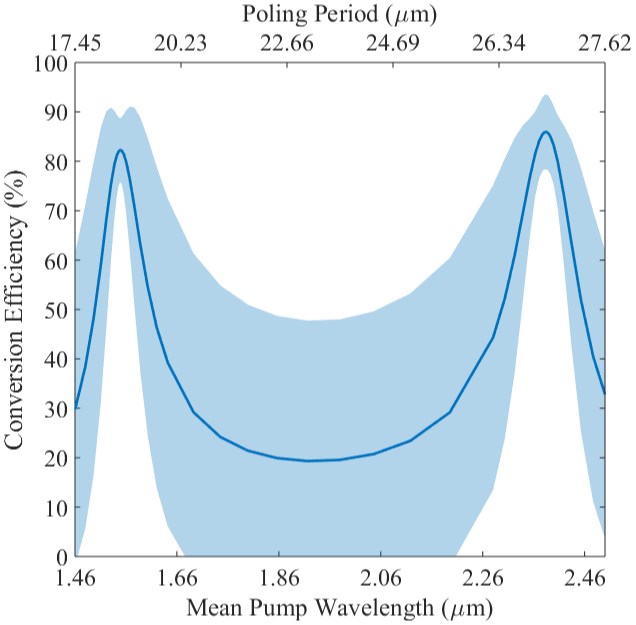}
    \caption{\label{fig3}Mean conversion efficiency over all C-band channels (spans 190.0 THz to 197.2 THz, minimum step of 100 GHz), against the average pump wavelength and the poling period. Shaded bars show one standard deviation in the efficiencies of all channels. It includes the effects of propagation loss, wrong pump interaction and pump SFG/SHG noise.}
\end{figure}

These results are summarized in Fig.~\ref{fig3} where the average conversion efficiency between all pairs of WDM channels, for a total of 5256 combinations, is plotted as a function of the mean pump wavelength $\sum_{n=1}^{5256}(\lambda_{\text{P1}_n}+\lambda_{\text{P2}_n})/10512$ and corresponding poling period. The shaded area in the figure represents one standard deviation around the average efficiency and is an indication of how broadband the process can be. These data are calculated numerically using the dispersion curve of bulk lithium niobate \cite{Edwards1984} for a 5 cm long device with an effective area of 25 {\textmu}m$^2$, pumps optimised according to eq. \ref{PowBal} and propagation losses of 0.1 dB/cm, which are standard for reverse proton exchanged waveguides. The system of coupled mode equations includes more modes than the original five to reflect wrong-pump interactions (+12 modes) and unwanted SFG/SHG interactions between the pumps (+7 modes).  Figure~\ref{fig3} shows that for pump frequencies close to the signals and targets, conversion efficiency is 82\% on average across the WDM spectrum, as $\delta_K$ remains small when the pumps are close to the signals. Ideal conversion ($\delta_K=0$) is not possible however, as this occurs when the pump frequencies are equal to the signal and target frequencies.
    
The most important result of this analysis is that because of the chromatic dispersion of the material, ideal conversion can be obtained with pumps near 2.38 {\textmu}m. This is possible because 2.38 {\textmu}m lies on the opposite side of an inflection in the dispersion curve of lithium niobate to 1.55 {\textmu}m, allowing $\delta_K$ to go to zero. In this case the conversion efficiency is 86\% on average across the whole WDM spectrum, as shown in Figure~\ref{fig3}. This improvement over the case of telecom pumps is due to the lower dispersion around 2.38 {\textmu}m, meaning $\delta_K$ remains smaller across the entire WDM spectrum. Additionally, the entire WDM spectrum is available for use as the pumps are well separated from the signals.

The drawback to using 2.38 {\textmu}m pumps is that the effect of interaction with the wrong pump is made worse.  The chromatic dispersion around $\omega_\text{SFG}$ is smaller than it is for 1.55 {\textmu}m pumps. This makes the average phase mismatch $\bar{K}$ of the unwanted process smaller in magnitude so the process is more efficient.  To illustrate, using the data calculated for Fig. \ref{fig3}, 2.1\% of the signal power is lost to crosstalk for a 200 GHz step using 1.55 {\textmu}m pumps.  This goes up to 7.8\% lost using 2.38 {\textmu}m pumps for the same 200 GHz step. The cost for mitigating this crosstalk is to fabricate longer devices, which provides a technical challenge, or to make the channels more broadly spaced. The loss to crosstalk drops below 1\% for frequency steps larger than 600 GHz.
		
\section{Experimental results}
The conversion protocol was used to perform a set of frequency conversion measurements using the set-up of Fig.~\ref{fig1}. 
Pump lasers in the C-band were used resulting in a non-optimal conversion as shown in Fig.~\ref{fig3}. 
This choice was made because tunable lasers around 2.38~{\textmu}m wavelength, which give higher conversion efficiency, were not available in our laboratory. 

The central element of the set-up is a nonlinear periodically poled waveguide in lithium niobate fabricated using the reverse proton exchange technique \cite{Lenzini2015,Kasture_JoO}. The device is 6~cm long, with a 5~cm poled region of period $\Lambda$= 16.02~{\textmu}m, and nominal propagation losses of 0.1~dB/cm. The device was also heated to $106^o$C to avoid photorefractive damage at high pump power and to tune the resonance frequencies of the nonlinear processes to be in line with the WDM channels.

\begin{figure}[h]
    \includegraphics[width=\columnwidth]{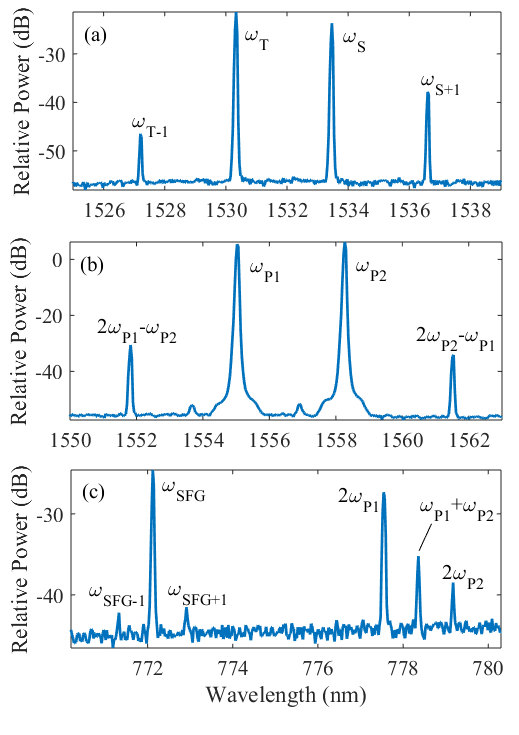}
    \caption{\label{fig4} Spectra from the output of the waveguide. \textbf{(a)} Spectrum around the signal and target wavelength. Two other peaks are present generated by the interaction of signal and idler with the wrong pumps. These peaks are at least 10dB smaller that the correct ones. \textbf{(b)} Pump spectral region. The other peaks present are generated by DFG between the pumps and their own sum frequencies and second harmonics.  \textbf{(c)} Spectra in the SFG region. Several parasitic nonlinear process can be seen, with wrong-pump peaks occurring on either side of $\omega_\text{SFG}$ and SHG/SFG between the pumps to the right.}
\end{figure}

To characterize the waveguide and estimate its conversion efficiency, we measured the second harmonic generation as a function of the pump wavelength and estimated an interaction length of 3.2~cm with an effective area of the SHG process of 42~{\textmu}m$^2$. Overall insertion losses of the device are 70\% which gives an estimated coupling loss of 66\%. See supplementary material for details on this measurement.

\begin{table*}
    \caption{\label{tab1}%
        Summary of the main results of the frequency conversion measurements. Wavelengths of signal, target and pumps, pump power levels used, measured efficiencies, and their theoretical estimations.}
    \begin{ruledtabular}
        \begin{tabular} {cccccccccc}
            $\lambda_\text{s}$ (nm) & $\lambda_\text{t}$ (nm) & $\lambda_\text{P1}$ (nm) & $\lambda_\text{P1}$ (nm) & $P_1$ (mW) & $P_2$ (mW) &$\eta^\text{exp}_c$ & $\eta^\text{exp}_d$ & $\eta^\text{th}_c$ & $\eta^\text{th}_d$\\
            \colrule
            1533.465	& 1531.898	& 1555.021	& 1556.636	& $131.1\pm0.9$	& $124.9\pm0.9$	& $0.19\pm0.01$	& $0.070\pm0.004$	& 0.403	& 0.123\\
            1533.465	& 1530.334	& 1555.021	& 1558.254	& $121.0\pm0.9$	& $127\pm1$		& $0.28\pm0.02$	& $0.16\pm0.01$		& 0.385	& 0.156\\
            1530.334	& 1535.036	& 1558.254	& 1553.409	& $136\pm1$		& $121\pm1$		& $0.31\pm0.02$	& $0.134\pm0.008$	& 0.403	& 0.107\\
            1535.036	& 1528.773	& 1553.409	& 1559.875	& $125\pm1$		& $125\pm1$		& $0.28\pm0.02$	& $0.162\pm0.007$	& 0.387	& 0.144\\
            1528.773	& 1536.609	& 1559.957	& 1551.881	& $124.0\pm0.9$	& $125.4\pm0.9$	& $0.26\pm0.02$	& $0.18\pm0.02$		& 0.385	& 0.146\\
        \end{tabular}
    \end{ruledtabular}
\end{table*}

The two pump lasers were combined into a single fiber using a commercial WDM module with channel spacing of 200~GHz. The pass band of this WDM module allowed up to $\pm 40$ GHz of tuning around the peak frequencies, and different frequency conversions were performed using different WDM channels. The signal beam was combined with the two pumps with a 90/10 fiber coupler with the pumps coupled into the 90\% arm in order to maximize the amount of pump power available for the experiment.

Five different frequency conversion experiments were performed with frequency shifts between signal and target ranging from 0.2 to 1~THz. During the experiments the output of the waveguide was collected with an achromatic lens and sent to a 50/50 beamsplitter. After the beamsplitter, light was sent to power meters and an optical spectrum analyzer (OSA) simultaneously. Data collection was automated and the pump frequencies were scanned in 5 GHz increments to find the peak conversion efficiency. Pump relative powers were balanced using traces from the OSA for each measurement while the signal power transmitted through the waveguide was always around 1~mW. Values of the wavelengths used, pump powers, and efficiencies measured and calculated are summarized in Tab.\ref{tab1}. 

Figure~\ref{fig4} shows the OSA traces of the frequency conversion from $\lambda_s$=1555.021~nm to $\lambda_t$=1558.254~nm. From these data we can see that several parasitic nonlinear effects are present. Figure~\ref{fig4}c shows that the correct SFG field as the highest peak but second harmonics from the two pumps as well as SFG between the two pumps are also present. The smaller peaks near the SFG one are generated by interaction with the wrong pumps as shown in Fig.~\ref{fig2}b. Figure~\ref{fig4}a shows the signal and target peaks and the cross talk with other two unwanted channel. In this particular case the power on the unwanted channels was 16.46 dB and 25.12 dB smaller than the power in the target channel. From these traces it is possible to obtain the relative intensities of all the frequency components, and when combined with the total power measured with the power meters absolute power measurements of each component can be inferred. See supplementary material for the full data set.

\begin{figure}[t]
    \includegraphics[width=\columnwidth]{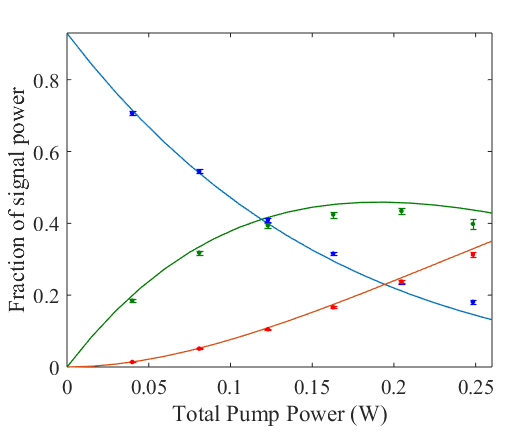}
    \caption{\label{fig5} Frequency conversion as a function of the total pump power for signal (blue), SFG (green), and target (red). Dots are experimental measurements while solid lines are from theoretical calculations. The pump powers for $P_1$ and $P_2$ are balanced according to Eq.~\ref{PowBal}.}
\end{figure} 

The conversion protocol was further tested by measuring the dependence of the converted power as a function of the combined pump powers, with the powers of $P_1$ and $P_2$ approximately balanced. Measurements up to a total pump power of 250~mW are shown in Fig.~\ref{fig5}.
The markers represent the relative power levels of signal, target, and SFG while the solid lines are calculated from the theory. From this trend it is estimated that maximum conversion requires 740~mW total pump power for a conversion efficiency $\eta_c=0.998$ and $\eta_d=1-1.02\times10^{-6}$. If the device was performing to the specified design, using the whole 5~cm of poling and with an effective area of 25~{\textmu}m$^2$, maximum conversion would only require 180~mW total pump power for a conversion efficiency $\eta_c=0.993$ and $\eta_d=1-1.22\times10^{-5}$.

\section{Conclusion}
By cascading $\chi^{(2)}$ nonlinear optical processes, we can achieve conversion between arbitrary frequencies and also overcome the speed bottleneck produced by a measure and retransmit protocol.
Our analysis of the role of phase mismatch has allowed us to improve upon other nonlinear optical schemes in terms of tuning bandwidth and efficiency.
In a device with a single fixed poling period, there is an optimal choice of pumps which minimises the average phase mismatch for any given conversion and maximises the overall efficiency.
Furthermore, in lithium niobate, using pumps near 2.38 {\textmu}m is both more efficient on average and leaves the entire C-band free from pumps and pump noise.

Our experimental results illustrate that this technique is quite feasible considering the maturity of our chosen platform, reverse proton exchanged waveguides in lithium niobate.
Regardless, our formulation of this frequency conversion technique applies to any $\chi^{(2)}$ nonlinear material or waveguide platform.
An alternative, such as etched waveguides in thin film lithium niobate, may provide the technology necessary to achieve longer interaction lengths more reliably.

\section{Acknowledgments}
This work was supported by the Australian Research Council (ARC) Centre of Excellence for Quantum Computation and Communication Technology (CE170100012), and the Griffith University Research Infrastructure Program. ML was supported by the Australian Research Council (ARC) Future Fellowship (FT180100055). PF and MV were supported by the Australian Government Research Training Program Scholarship. This work was performed in part at the Queensland node of the Australian National Fabrication Facility, a company established under the National Collaborative Research Infrastructure Strategy to provide nano- and microfabrication facilities for Australia's researchers. 

\bibliography{cascadedFCBib}

\newpage
\onecolumngrid

\section*{Supplementary Material for ``An integrated optical device for frequency conversion across the full telecom C-band spectrum''}
\section*{Mathematics}
    We begin with electric fields present in a z-propagating waveguide of the form
    \begin{equation}
    E_j = \frac{1}{2}\rho_n(x,y)\left(A_j e^{i(\omega_j t - k_j z)} + \text{c.c.} \right) ,
    \end{equation}
    for each $j$ being the frequencies of interest, Signal (S), Target (T), Pump1 (P1), Pump2 (P2) and SFG. Here, $A_j$ is the complex amplitude of the mode, $\omega_j$ is its angular frequency, $k_j$ is its wavenumber (including effects of waveguide dispersion, often denoted as $\beta$), and $\rho(x,y)$ is its cross-sectional mode profile.  Following a standard derivation of 2nd order nonlinear optics, using slowly varying envelope and undepleted pump approximations ($A'_\text{P1}$ \& $A'_\text{P2}$ constant) we reach the system of coupled differential equations,
    \begin{equation}
    \frac{dA'_\text{S}}{dz} = -iJ_\text{S}S_\text{S}A'^*_\text{P1}A'_\text{SFG}e^{-i\Delta k_1z}\ ,
    \end{equation}
    \begin{equation}
    \frac{dA'_\text{T}}{dz} = -iJ_\text{T}S_\text{T}A'^*_\text{P2}A'_\text{SFG}e^{-i\Delta k_2z}\ ,
    \end{equation}
    \begin{equation}
    \frac{dA'_\text{SFG}}{dz} = -iJ_\text{SFG}\left[S_\text{SFGa}A'_\text{P1}A'_\text{S}e^{i\Delta k_1z}+S_\text{SFGb}A'_\text{P2}A'_\text{T}e^{i\Delta k_2z}\right]\ .
    \end{equation}
    In this formulation, $\Delta k_1 = k_\text{SFG}-k_\text{S}-k_\text{P1}$ and $\Delta k_2 = k_\text{SFG}-k_\text{T}-k_\text{P2}$ are the phase mismatches of the SFG and DFG processes respectively.  We have also substituted,
    \begin{equation}
    A'_j = A_j\sqrt{\frac{\epsilon_0c}{2 n_{\text{eff},j}}\iint n^2_j \rho^2_j dxdy} ,
    \end{equation}
    with $\epsilon_0$ as the permittivity of free space, $c$ as the speed of light, $n$ the refractive index and $n_{eff}$ the effective index of the mode. This is done so that $\left|A'_j\right|^2 = P_j$, the power carried by the waveguide at that frequency.
    The parameters $S_j$ take into account the overlap between the interacting modes and are given by,
    \begin{align}
    S_\text{S} &= \frac{\sqrt{n_\text{eff,P1} n_\text{eff,SFG}}}{n_\text{eff,S}\sqrt{n_\text{eff,S}}}
      \frac{\iint \rho_\text{S}\rho_\text{P1}\rho_\text{SFG}dxdy}{\iint \rho^2_\text{S} dxdy}
      \sqrt{\frac{\iint n_\text{S}^2 \rho_\text{S}^2 dxdy}{\iint n_\text{P1}^2 \rho_\text{P1}^2 dxdy \iint n_\text{SFG}^2 \rho_\text{SFG}^2 dxdy}} , \\
    S_\text{P1} &= \frac{\sqrt{n_\text{eff,SFG} n_\text{eff,S}}}{n_\text{eff,P1}\sqrt{n_\text{eff,P1}}}
      \frac{\iint \rho_\text{P1}\rho_\text{SFG}\rho_\text{S}dxdy}{\iint \rho^2_\text{P1} dxdy}
      \sqrt{\frac{\iint n_\text{P1}^2 \rho_\text{P1}^2 dxdy}{\iint n_\text{SFG}^2 \rho_\text{SFG}^2 dxdy \iint n_\text{S}^2 \rho_\text{S}^2 dxdy}} , \\
    S_\text{SFGa} &= \frac{\sqrt{n_\text{eff,S} n_\text{eff,P1}}}{n_\text{eff,SFG}\sqrt{n_\text{eff,SFG}}}
      \frac{\iint \rho_\text{SFG}\rho_\text{S}\rho_\text{P1}dxdy}{\iint \rho^2_\text{SFG} dxdy}
      \sqrt{\frac{\iint n_\text{SFG}^2 \rho_\text{SFG}^2 dxdy}{\iint n_\text{S}^2 \rho_\text{S}^2 dxdy \iint n_\text{P1}^2 \rho_\text{P1}^2 dxdy}} ,
    \end{align}
    with a corresponding set for $S_\text{T}$, $S_\text{P2}$ and $S_\text{SFGb}$. These can be converted into effective interaction areas for the processes related to the spatial overlap of the modes,
    \begin{align}
    A_\text{eff,SFG} &= \left[ n_\text{eff,S} n_\text{eff,P1} n_\text{eff,SFG} S_S S_{SFGa} \right]^{-1}\\
    A_\text{eff,DFG} &= \left[ n_\text{eff,T} n_\text{eff,P2} n_\text{eff,SFG} S_T S_{SFGb} \right]^{-1}
    \end{align}
    Finally,
    \begin{equation}
    J_j = \frac{\omega_j \chi^{(2)}_\text{eff}}{c}\sqrt{\frac{2}{\epsilon_0 c}} ,
    \end{equation}
    collects all of the constants into a single term, with $\chi^{(2)}_\text{eff}$ as the effective nonlinear coefficient in the waveguide.
    
    We introduce the quantities of Average Phase Mismatch ($K$) and Difference in Phase Mismatch ($\delta_\kappa$) such that,
    \begin{equation}
    \Delta k_1 = K - \nicefrac{\delta_\kappa}{2} \quad , \quad \Delta k_2 = K + \nicefrac{\delta_\kappa}{2}.
    \end{equation}
    Using solutions of the form,
    \begin{equation}
    A'_\text{S} = \bar{A}'_\text{S} e^{i\frac{\delta_\kappa}{2}z} \quad , \quad A'_\text{T} = \bar{A}'_\text{T} e^{-i\frac{\delta_\kappa}{2}z} \quad , \quad A'_\text{SFG} = \bar{A}'_\text{SFG} e^{iKz}
    \end{equation}
    results in the linear system,
    \begin{equation}
      \frac{d}{dz}
      \begin{bmatrix}
        \bar{A}'_\text{S} \\
        \bar{A}'_\text{T} \\
        \bar{A}'_\text{SFG}
      \end{bmatrix}
      = 
      \begin{bmatrix}
        -i\frac{\delta_\kappa}{2} & 0 & -iJ_\text{S}S_\text{S}A'^*_\text{P1} \\
        0 & i\frac{\delta_\kappa}{2} & -iJ_\text{T}S_\text{T}A'^*_\text{P2} \\
        -iJ_\text{SFG}S_\text{SFGa}A'_\text{P1} & -iJ_\text{SFG}S_\text{SFGb}A'_\text{P2} & -iK
      \end{bmatrix}
      \begin{bmatrix}
      \bar{A}'_\text{S} \\
      \bar{A}'_\text{T} \\
      \bar{A}'_\text{SFG}
      \end{bmatrix}
      = M\mathbf{\bar{A}}
    \end{equation}
    which has the characteristic equation,
    \begin{equation}
    \lambda^3+iK\lambda^2+\left(\frac{\delta^2_K}{4} + Q\right)\lambda+i\frac{\delta_\kappa}{2}\left(K\frac{\delta_\kappa}{2}-P\right) = 0
    \end{equation}
    where,
    \begin{align}
    Q = J_\text{S}J_\text{SFG}S_\text{S}S_\text{SFGa}P_\text{P1} + J_\text{T}J_\text{SFG}S_\text{T}S_\text{SFGb}P_\text{P2}\ , \\
    P = J_\text{S}J_\text{SFG}S_\text{S}S_\text{SFGa}P_\text{P1} - J_\text{T}J_\text{SFG}S_\text{T}S_\text{SFGb}P_\text{P2}\ .
    \end{align}
    
    The general solution is
    \begin{equation}
    \mathbf{\bar{A}} = \mathbf{C}_1e^{\lambda_1z} + \mathbf{C}_2e^{\lambda_2z} + \mathbf{C}_3e^{\lambda_3z}
    \end{equation}
    where
    \begin{align}
    \mathbf{C}_1 = \frac{1}{(\lambda_2-\lambda_1)(\lambda_1-\lambda_3)} \left[-\lambda_2\lambda_3\mathbf{\bar{A}}(0)+(\lambda_2+\lambda_3)M\mathbf{\bar{A}}(0)-M^2\mathbf{\bar{A}}(0)\right] \\
    \mathbf{C}_2 = \frac{1}{(\lambda_3-\lambda_2)(\lambda_2-\lambda_1)} \left[-\lambda_1\lambda_3\mathbf{\bar{A}}(0)+(\lambda_1+\lambda_3)M\mathbf{\bar{A}}(0)-M^2\mathbf{\bar{A}}(0)\right] \\
    \mathbf{C}_3 = \frac{1}{(\lambda_3-\lambda_2)(\lambda_1-\lambda_3)} \left[-\lambda_1\lambda_2\mathbf{\bar{A}}(0)+(\lambda_1+\lambda_2)M\mathbf{\bar{A}}(0)-M^2\mathbf{\bar{A}}(0)\right]
    \end{align}
    
    In the case where we set the average phase mismatch to equal the poling period of the device by choosing the correct pump frequencies $(K=0)$ and we balance the pump powers $(P=0)$, the characteristic equation simplifies and we get eigenvalues,
    \begin{equation}
    \lambda_1 = 0 \quad , \quad \lambda_{2,3} = \pm \frac{i}{2}\sqrt{\delta^2_\kappa + 4Q}\ .
    \end{equation}
    When only the signal mode and pumps have power initially $(A'_\text{T}(0) = A'_\text{SFG}(0) = 0)$ the field amplitudes are,
    \begin{equation}
    A'_\text{S}(z) = \frac{A'_\text{S}(0)}{\delta_\kappa^2+4Q}\left\{ 2Q + (2Q+\delta_\kappa^2)\cos\left(\frac{z}{2}\sqrt{\delta_\kappa^2+4Q}\right) - i\delta_\kappa\sqrt{\delta_\kappa^2+4Q}\sin\left(\frac{z}{2}\sqrt{\delta_\kappa^2+4Q}\right) \right\}e^{i\frac{\delta_\kappa}{2}z}
    \end{equation}
    \begin{equation}
    A'_\text{T}(z) = \frac{-2QA'_\text{S}(0)}{\delta_\kappa^2 + 4Q}\sqrt{\frac{J_\text{T}S_\text{T}S_\text{SFGa}}{J_\text{S}S_\text{S}S_\text{SFGb}}}\left\{ 1 -  \cos\left(\frac{z}{2}\sqrt{\delta_\kappa^2+4Q}\right)\right\}e^{-i\frac{\delta_\kappa}{2}z}
    \end{equation}
    \begin{equation}
    A'_\text{SFG}(z) = \frac{-\sqrt{2Q}A'_\text{S}(0)}{\delta_\kappa^2 + 4Q}\sqrt{\frac{J_\text{SFG}S_\text{SFGa}}{J_\text{S}S_\text{S}}} \left\{ \delta_\kappa -  \delta_\kappa \cos\left( \frac{z}{2}\sqrt{\delta_\kappa^2 + 4Q} \right) 
     - i \sqrt{\delta_\kappa^2 + 4Q} \sin\left(\frac{z}{2}\sqrt{\delta_\kappa^2+4Q}\right) \right\}
    \end{equation}
    Since $\left|A'_j\right|^2 = P_j$, it is trivial to change these expressions into conversion and depletion efficiencies at $z=L$,
    \begin{equation}
    \eta_c = \frac{P_T}{P_S(0)}\frac{J_SS_SS_{SFGb}}{J_TS_TS_{SFGa}} = \frac{Q^2L^4}{16}\text{sinc}^4\left( \frac{L}{4}\sqrt{\delta_\kappa^2+4Q} \right)
    \end{equation}
    \begin{equation}
    \eta_d = 1 - \frac{P_S}{P_S(0)} = \frac{QL^2}{2}\text{sinc}^2\left( \frac{L}{4}\sqrt{\delta_\kappa^2+4Q} \right) - \frac{Q^2L^4}{16}\text{sinc}^4\left( \frac{L}{4}\sqrt{\delta_\kappa^2+4Q} \right)
    \end{equation}
  
  \section*{Estimating efficiency from SHG}
\begin{figure}[h]
        \includegraphics[width=0.5\columnwidth]{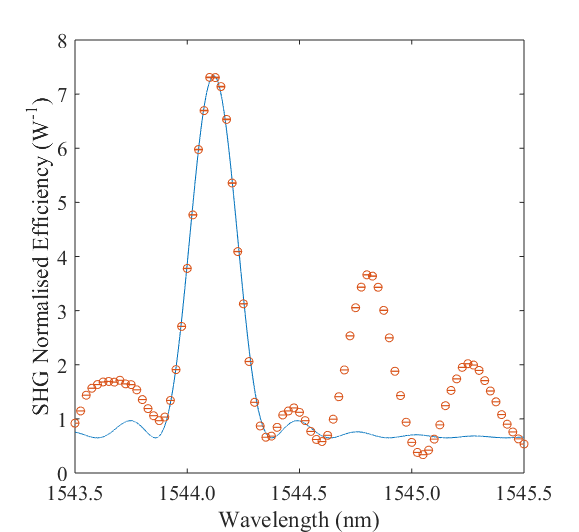}
        \caption{\label{figSHG} Measured SHG efficiency (circles) with superimposed fit (solid line) according to eq. \ref{etaSHG}.}
    \end{figure}

    The non-uniformity of the waveguide led to a multi-peaked SHG efficiency curve as shown in figure \ref{figSHG}.  The interaction length and effective mode area of the waveguide were estimated by fitting a vertically-offset sinc$^2$ curve to the tallest peak in this efficiency data, according to the equation,
    \begin{equation} \label{etaSHG}
        \eta_\text{SHG} = \frac{32\chi^{(2)2}_\text{eff}L_\text{int}^2}{\epsilon_0c\lambda^2 n^2_{\text{eff},\lambda} n_{\text{eff},\lambda/2} A_\text{eff}}
        \text{sinc}^2\left( \frac{2\pi L_\text{int}}{\lambda} \left[ n_{\text{eff},\lambda} - n_{\text{eff},\lambda/2} - \frac{\lambda}{\lambda_{0}} \left( n_{\text{eff},\lambda_0} - n_{\text{eff},\lambda_0/2} \right) \right] \right) + v
    \end{equation}
    where $L_\text{int}$ is the interaction length, $A_\text{eff}$ is the effective area of the waveguide, $\lambda_0$ is the central wavelength and $v$ is the vertical offset.
    The effective $\chi^{(2)}$ used for this calculation and all simulations and theoretical plots was $25\times10^{-6}${\textmu}m/V. The factor of $2/\pi$ for 1st order quasi-phase matching is included within equation \ref{etaSHG}.  The peak wavelength was 1544.11 nm, vertical offset was 0.65/W, effective length was $3.2\pm0.2$cm and the effective area was $42\pm2${\textmu}m$^2$, with uncertainties given by confidence bounds of the fit.
    \newpage
    
  \section*{Full data set}
  Complete set of data for all the conversion experiments reported in Table 1 of the manuscript. These plots are equivalent to Figure 5 in the manuscript which corresponds to the conversion between 1533.465 and 1530.334 nm.
  
  \begin{figure}[h]
      \includegraphics[width=0.75\columnwidth]{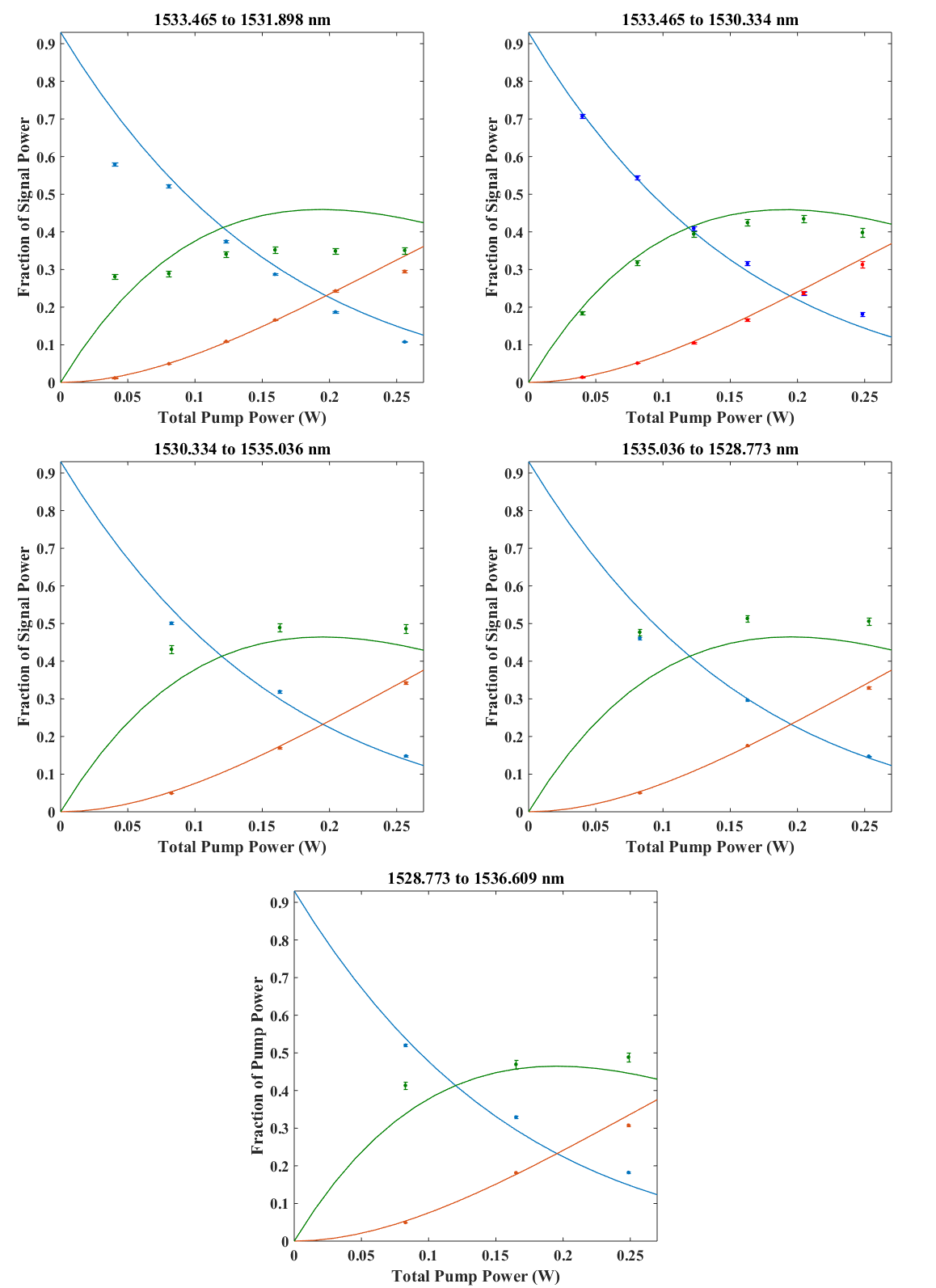}
      \caption{\label{figAllData} Complete data set for all 5 conversions, plotting frequency conversion as a function of total pump power for signal (blue), SFG (green) and target(red).  Dots are experimental measurements while solid lines are from theoretical calculations.}
  \end{figure}
\end{document}